\documentclass[amsmath,showpacs,twocolumn,aps,prl]{revtex4}
\usepackage{bm}
\usepackage{graphics}
\begin{document}

\title{Charge response function and a novel plasmon mode in graphene}

\author{S.~Gangadharaiah, A.~M.~Farid, and E.~G.~Mishchenko}
\affiliation{Department of Physics, University of Utah, Salt Lake
City, Utah 84112, USA}
\begin{abstract}

Polarizability of non-interacting 2D Dirac
electrons has a $1/\sqrt{qv-\omega}$ singularity at the boundary of
electron-hole excitations. The screening of this singularity by long-range electron-electron
interactions is usually treated within the random phase
approximation. The latter is exact only in the limit of $N\to
\infty$, where $N$ is the ``color'' degeneracy. We find that the
ladder-type vertex corrections become crucial close to the threshold as
the ratio of the $n$-th order ladder term to
the same order RPA contribution is
 $\ln^n|qv-\omega|/N^n$.
  We perform analytical summation of the infinite series of
ladder diagrams which describe excitonic effect. Beyond the
threshold,  $qv>\omega$, the real part of the polarization operator
is found to be positive leading to the appearance of a strong and
narrow plasmon resonance.

\end{abstract}


\maketitle

{\it Introduction.} Many properties of interacting two-dimensional
electrons with linear Dirac spectrum $\epsilon=\pm vp$, found in
 graphene monolayers, differ sharply from those with the parabolic
spectrum present in conventional semiconductor heterostructures
\cite{Wil,cn}. Vanishing density of states at the Dirac point and less
effective screening of Coulomb interaction lead to electronic
properties of graphene being qualitatively different.

One of the
distinct signatures of non-interacting 2D electrons in graphene,
caused by the absence of spectrum curvature, is a divergent behavior
of the polarization operator (charge susceptibility)
at the threshold for the excitation of electron-hole pairs
\cite{Shung,Gonzales1,Gonzales2,Chubukov, Khveshchenko1},
$\Pi^{(0)}(\omega,q)=-Nq^2/16\sqrt{q^2v^2-\omega^2}$, where $N$ is
the degree of degeneracy (in graphene $N=4$ in the absence of
magnetic field).  
The effects of electron-electron
interaction on the polarization operator are customarily accounted
for by the random phase approximation (RPA)
\cite{Gonzales2,Khveshchenko2,Sarma,Barlas}, which sums the
infinite series of electron loops,
\begin{equation}
 \Pi^{-1}_{RPA}=(\Pi^{(0)})^{-1}-V_q,\label{rpa}
\end{equation}
where $V_q=2\pi e^2/q$, dielectric constant of the substrate being
incorporated into the charge $e^2$. In particular, RPA predicts that
the imaginary part of the polarization operator (which determines
absorption in the system and the density-density correlation
function) instead of diverging actually vanishes at the threshold
according to, $-\Pi''_{RPA}\propto g^{-2}\sqrt{\omega-qv}$, where
$g=e^2/v$ is the dimensionless interaction constant. 

A notable property of 2D electrons, well captured by RPA in a
conventional  Fermi liquid, is the existence of a low-frequency
collective mode of charge  oscillations whose spectrum is,
$\omega_q^2=2e^2E_F q$, where $E_F$ is the Fermi energy. In undoped
graphene, $E_F=0$, the plasmon mode is thus {\it absent} {\it within
RPA} (finite doping or temperature can lead to the usual RPA
plasmons \cite{E_plasmon,T_plasmon}). Mathematically this comes from
the fact that the real part of $\Pi^{(0)}$ is {\it negative} for
$qv>\omega$ (i.e. beyond the domain of electron-hole excitations).
In the present paper we demonstrate that this conclusion is an
artifact of the approximation (\ref{rpa}) and that a collective mode
does {\it exist} in undoped graphene. This mode describes charge
fluctuations in a system of electron-hole pairs {\it interacting} in
a final state (``excitonic'' effect), rather than {\it free} pairs,
as implied by Eq.~(\ref{rpa}).

We begin with noting that RPA is much less justified in case
of undoped graphene than in the case of conventional parabolic
spectrum, where it is formally valid for momenta less than the
inverse screening length, $q\ll \kappa$. In undoped graphene the
vanishing density of carriers ensures that $\kappa \to 0$. In
particular, direct calculation of the electron self-energy to the
second order in interaction demonstrates that non-RPA contributions
are generally smaller than the RPA terms only by virtue of $1/N$
\cite{Mishchenko}. The RPA is exact only in the formal limit of
$N\to \infty$, since each RPA term corresponds to the largest number of
loops possible in each order of the perturbation theory.

Below we analyze consecutive orders of the perturbation series in
the bare interaction $V_q=2\pi e^2/q$ and compare RPA diagrams to
the ladder corrections. We find that the ratio of the $n$-th ladder
term to the corresponding RPA  contribution of the same order is
$N^{-n}\ln^n{(\frac{qv}{|qv-\omega|})}$ and large close to the
threshold. After extracting the leading singular terms in each
order, we calculate the infinite ladder series and obtain a
non-perturbative expression for the polarization operator near the
threshold. Summation of the leading divergencies {\it is possible}
once we realize that for ${qv}/{|qv-\omega|} \gg 1$ the main
contribution arises from the processes that involve
quasi-one-dimensional motion of interacting particles along the
direction of the external momentum ${\vec  q}$. In general, when $N
\sim \ln{(\frac{qv}{|qv-\omega|})}$, the perturbation expansion of
the polarization operator receives similar contributions from  both
the  ladder  and   RPA terms. Thus, the polarization operator is
composed of {\it all possible} combinations of ladders and loops.



{\it Non-interacting electrons.}  To demonstrate our method and to
introduce some notations it is instructive to begin with reproducing
the zeroth-order polarization operator. Close to the resonance, when
$|\omega-qv| \ll qv$, only those transitions are important that have
momenta ${\bf p}$ and ${\bf q}$ directed almost along the same line.
More precisely, the electron is taken from the state in the lower
cone with the momentum ${\bf p}$ almost antiparallel to ${\bf q}$
and placed in the upper cone in the state with momentum ${\bf
p}+{\bf q}$ which is almost parallel to ${\bf q}$. Obviously, this
is possible only as long as $p< q$. The relevance of these
particular transitions is understood from the form of the
polarization operator,
 \begin{equation}
 \label{pi0}
\Pi^{(0)}(\omega,{q})=-iN\text{Tr}\sum_{\bf p}\int
\frac{d\epsilon}{2\pi} \hat G_{\epsilon+\omega,{\bf p+q}} \hat
G_{\epsilon {\bf p}},
\end{equation}
where the electron Green's function consists of the contributions
from both cones $\beta=\pm 1$,
\begin{equation}
 \label{subband}
 \hat G_{\epsilon {\bf p}}=
 \frac{1}{2} \sum_{\beta=\pm 1} \frac{1+\beta \hat \sigma_{\bf  p}}{\epsilon-\beta
 v p+i\beta \eta},
 \end{equation}
 where $\hat \sigma_{\bf  p} =\hat {\bm \sigma}\cdot {\bf p}/p$ is
 the projection of the pseudospin Pauli matrix onto the direction of
 electron momentum.
 The energy integral in Eq.~(\ref{pi0}) yields,
 \begin{equation}
 \label{pi01}
\Pi^{(0)}(\omega,{q})=\frac{N}{4}\sum_\beta \sum_{\bf
p}\frac{\text{Tr}(1+\beta\hat\sigma_{\bf
p+q})(1-\beta\hat\sigma_{\bf p})}{\omega-\beta v(p+|{\bf p}+{\bf
q}|)+i\beta\eta},
\end{equation}
we observe that i) for positive $\omega$ the singular denominator
appears for $\beta=+1$, ii) the trace operation in the numerator
imposes that ${\bf p}$ and ${\bf p}+{\bf q}$ are antiparallel (for
the parallel configuration the trace vanishes).
Close to the threshold ($\omega= qv$) collinear processes become dominant
allowing us to approximate $|{\bf p}+{\bf q}| \approx q-p+\frac{pq\theta^2}{2(q-p)}$, where
$\theta$ denotes the angle between ${\bf p}$ and $-{\bf q}$ and
$\text{Tr}(1+\hat\sigma_{\bf p+q})(1-\hat\sigma_{\bf p})\approx 4$,
thus we  obtain from Eq.~(\ref{pi01}), 
\begin{equation}
\label{im_p_approx} \Pi^{(0)}(\omega,{q})= N \int\limits_0^q
\frac{pdp}{(2\pi)^2} \int\limits_{-\infty}^{\infty} \frac{
d\theta}{\omega -qv-v\frac{pq\theta^2}{2(q-p)}+i\eta}.
\end{equation}
Upon calculating the integrals we recover
$\Pi^{(0)}(\omega,{q})=-\frac{N}{16\sqrt{2}v}
q^{3/2}/\sqrt{qv-\omega}$. Note that, as expected, the relevant
angles are small $ \theta \sim \sqrt{|1-\omega/qv|} \ll 1$.
\begin{figure}[h]
\resizebox{.40\textwidth}{!}{\includegraphics{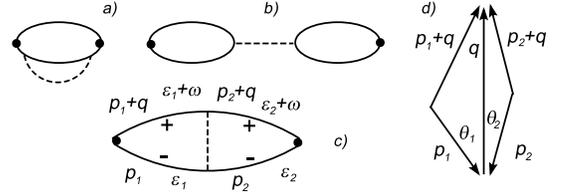}}
\caption{First-order interaction correction: self-energy correction a),
sub-leading RPA correction b), leading vertex correction c). Figure d) illustrates
the origin of the singular behavior, $\sim
(qv-\omega)^{-1}\ln{|qv-\omega|}$ from almost collinear electron
propagation. Both states ${\bf p}_1$, ${\bf p}_2$ belong to the
lower ($-$) cone, while ${\bf p}_1+{\bf q}$, ${\bf p}_2+{\bf q}$
correspond to the upper ($+$) cone.}
\end{figure}

{\it First-order correction.} To the first order in the
electron-electron interaction there are three diagrams, Fig.~1. Of
these diagrams, the self-energy correction, Fig.~1a), is the most
singular near the threshold,
\begin{equation}
\label{P_self-energy} \Pi^{(1)}_{SE}(\omega,q)= -\frac{N
g\sqrt{v}}
{16^2\sqrt{2}}\frac{q^{5/2}}{(qv-\omega)^{3/2}}\ln{({\cal K}/q)},
\end{equation}
where ${\cal K}$ is the upper momentum cut-off. This singularity
indicates that the infinite summation of the self-energy diagrams
have to be performed first, thus yielding the renormalization of the
electron spectrum. To the lowest order in interaction this gives $v_p= v[1 +\frac{g}{4}\ln{({\cal K}/p)}]$ \cite{Gonzales1}. In the case of a conventional Fermi liquid the corresponding velocity renormalization is trivial. In case of a linear spectrum considered here, however, logarithmic terms lead to a slight curvature of electron spectrum which can smear the threshold singularities. In what follows we assume the renormalized velocity but neglect this smearing by taking the velocity at a typical momentum, determined by the external momentum $v\equiv v_q$.

The next term, 1b), is the RPA-correction which can be readily written as,
\begin{equation}
\label{P_RPA} \Pi^{(1)}_{RPA}(\omega,q)= \frac{\pi N^2
g}{16^2}\frac{q^2}{qv-\omega}.
\end{equation}

It can  now be verified that the vertex correction, Fig.~1c), yields a
contribution which is opposite in sign and more singular than Eq.~(\ref{P_RPA}) as a result of the  long-range character of Coulomb interaction. The energy integrations
are performed independently in each half of the diagram, resulting
in
\begin{eqnarray}
\Pi_V^{(1)} (\omega, q)&=&-\frac{N}{16}\text{Tr} \sum_{{\bf p}_1 {\bf
p}_2} \sum_{\alpha \beta} \frac{\beta(1-\beta \sigma_{{\bf
p}_1+q})(1+\beta \sigma_{{\bf p}_1})}{\omega+\beta
v(p_1+|{\bf p}_1+{\bf q}|)}\nonumber\\
&&\times  \frac{\alpha(1+\alpha \sigma_{{\bf p}_2})(1-\alpha
\sigma_{{\bf p}_2+{\bf q}})}{\omega+\alpha v(p_2+|{\bf p}_2+{\bf
q}|)}V_{{\bf p}_1-{\bf p}_2}.
\end{eqnarray}
To extract the leading contribution we note that it comes from
almost collinear processes with $\alpha=\beta=+1$ and both ${\bf
p}+{\bf q}$, ${\bf p'}+{\bf q}$ are almost parallel to ${\bf q}$ and
almost antiparallel to both ${\bf p}$, ${\bf p'}$. Using the
small-angle expansion introduced in the previous section, and
approximating $V_{\bf p-\bf p'} \approx  e^2/|p-p'|$ we obtain
\begin{equation}
\label{pi1} \Pi^{(1)}_V (\omega, q)=-\frac{N e^2}{(2\pi)^4}
\int\frac{1}{|p_1-p_2|} \prod_{i=1}^2 \frac{p_idp_i
d\theta_i}{\omega-qv -v \frac{p_iq\theta_i^2}{2(q-p_i)}}.
\end{equation}
The integrals over the two angles are independent and yield the same
resonant denominator as in Eq.~(\ref{P_RPA}). However, the integrals
over the magnitude of electron momenta contain a logarithmic
divergence which is due to the long-range nature of Coulomb
interaction. To cut-off this divergence we note that,  $|{\bf
p}_1-{\bf p}_2| \approx |p_1-p_2|
+\frac{p_1p_2\theta_{12}^2}{2|p_1-p_2|}$, where $\theta_{12}$ is the
angle between two momenta. Recalling that relevant angles
$\theta^2_{12}\sim |qv/\omega -1|$ and that typical $p_i \sim q$, we
observe that small-angle expansion fails for $|p_1-p_2| \sim q\sqrt{
|qv/\omega -1|}$,  which should be used as the lower cut-off. As a
result we obtain,
\begin{equation}
\label{P_vertex} \Pi^{(1)}_V(\omega, q)=\frac{gN}{24\pi}
\frac{q^2}{\omega-qv}\ln{\left(\frac{qv}{|\omega-qv|}\right)}.
\end{equation}
We, therefore, find that RPA loops exceed the vertex terms only when
$N \gg \ln{(\frac{qv}{|\omega-qv|})}$. In the opposite limit, close
to the resonance, $N \ll \ln{(\frac{qv}{|\omega-qv|})}$ the ladder
diagrams dominate. We now turn to the summation of the ladders
series.

{\it Summation of the infinite ladder.} In the second order in
electron-electron interaction the most dominant contribution again
comes from the ladder diagram, Fig.~2a), with three pairs of
electron lines each yielding $1/\sqrt{qv-\omega}$-singularity, and
two Coulomb interactions providing two additional powers of the
logarithm. The other diagrams are less singular at $\omega =qv$.
Diagram 2b) requires that at least one of the interaction lines
carries large momentum, $\sim q$, while both momenta in the
ladder-type diagram 2a) are small. Exchange-type contributions,
2c)-e), though providing the main power-law singularity, involve
transferred momenta which are $\sim q$ and lack the additional
logarithms (while also lacking extra factors $N$).
\begin{figure}[h]
\resizebox{.37\textwidth}{!}{\includegraphics{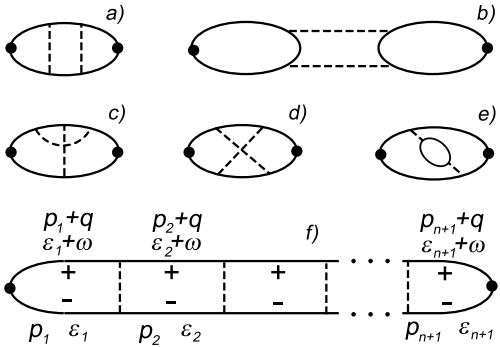}}
\caption{Leading, a), and subleading, b)-e), non-RPA diagrams of the
second order. The ladder diagram f) explains the origin of the most
singular, $(qv-\omega)^{-n}\ln^n{|qv-\omega|}$ contribution of the
arbitrary order $n$. All states propagating along the upper/lower
part of the ladder belong to the upper/lower cone.}
\end{figure}
Repeating the arguments from the preceding section  we realize that to each order in interaction
the leading contribution comes from the ladder diagram, Fig.~2f),
\begin{eqnarray*}
\Pi^{(n)}_V(\omega,q)=-N\text{Tr} \int \Bigl[\prod_{i=1}^{n+1}
\frac{d\epsilon_i}{2\pi} d{\bf p}_i \Bigr] V_{{\bf p}_1-{\bf
p}_2}... V_{{\bf p}_n-{ \bf p}_{n+1}} \nonumber \\ \times i^{n+1}
\hat G_{\epsilon_1+\omega {\bf p}_1+{\bf q}}... \hat
G_{\epsilon_{n+1}+\omega {\bf p}_{n+1}+{\bf q}} \hat
G_{\epsilon_{n+1} {\bf p}_{n+1}}...\hat G_{\epsilon_1 {\bf p}_1}.
\end{eqnarray*}
All the momenta are almost collinear, with those propagating along
the lower part of the ladder, ${\bf p}_i$, being antiparallel to
external wavevector ${\bf q}$ and to all the momenta ${\bf p}_i+{\bf
q}$ along the upper part of the ladder. Energy integrations are
independent in each step and result in the singularity to be the
strongest when all the propagators in upper/lower parts of the ladder
belong to the upper(+)/lower(-) cones. The subsequent trace
operation is easily performed as
\begin{eqnarray*}
\text{Tr} ~(1-\sigma_{{\bf p}_1})...(1-\sigma_{{\bf
p}_{n+1}})~(1+\sigma_{{\bf p}_{n+1}+{\bf q}})...\nonumber\\
... \times (1+ \sigma_{{\bf p}_1+{\bf q}}) \approx 4^{n+1},
\end{eqnarray*}
yielding the $n$-th order contribution to the polarization operator
in the form
\begin{eqnarray}
\label{pin}
 \Pi^{(n)}_V(\omega,q)&=& (-1)^{n} N \int V_{{
p}_1-{p}_2}... V_{{ p}_n-{ p}_{n+1}} \nonumber\\ && \times
\prod_{i=1}^{n+1}\frac{d{\bf p}_i}{\omega-v (p_i+|{\bf p}_i+{\bf
q}|)+i \eta}.
\end{eqnarray}
Integrating over the angles and then evaluating the integrals over
the absolute values of momenta in the leading logarithmic
approximation we obtain,
\begin{equation}
\Pi^{(n)}(\omega,q)=-\frac{1}{8\sqrt{\pi}} \frac{q^2}{
\sqrt{2qv(qv-\omega)}}
\frac{\Gamma\left(\frac{3+n}{2}\right)}{\Gamma\left(\frac{4+n}{2}\right)}x^n,
\end{equation}
where we introduced the following dimensionless variable, $
x=\frac{g}{2\sqrt{2}} \sqrt{\frac{qv}{qv-\omega}} \ln  (\frac{
qv}{|qv-\omega|} )$. The summation of the infinite series can now be
readily performed,
\begin{equation}
\label{pi_exact} \Pi (\omega,q)=-Nq
\frac{1+\frac{2}{\pi}\arcsin{x}-(1+\frac{2}{\pi}x)\sqrt{1-x^2}}{4 v
g \ln( \frac{ qv}{|qv-\omega|})~ x \sqrt{1-x^2}}.
\end{equation}

The above result (\ref{pi_exact}) for the polarization ladder is
sufficient to calculate the polarization operator in a more general
case when $N \sim \ln  (\frac{ qv}{|qv-\omega|}) \gg 1$. In this
regime {\it both}  vertex and loop diagrams have to be summed
simultaneously. Such summation can be performed by realizing that it
yields a geometric series similar to the usual RPA one (\ref{rpa})
but with the bare polarization operator $\Pi^{(0)}$ replaced by the
polarization ladder $\Pi_V$ given by Eq.~(\ref{pi_exact}),
\begin{eqnarray}
\label{new_RPA}
{\cal P} (\omega,q) = \frac{\Pi_V (\omega,q)}{1- V(q) \Pi_V(\omega,q)},
\end{eqnarray}

\begin{figure}[h]
\resizebox{.37\textwidth}{!}{\includegraphics{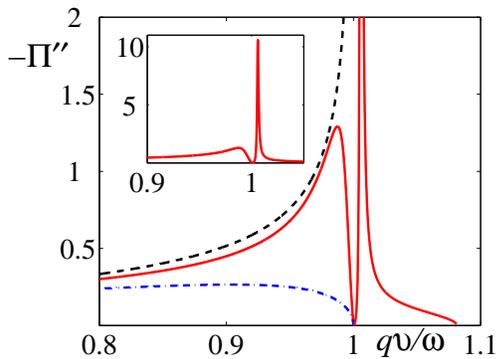}}
\caption{(Color online) Imaginary part of the polarization operator
in units of $q/4$ for $N=4$ and $g=0.3$: non-interacting value
$\Pi''_0$ (black dashed line), RPA value $\Pi''_{RPA}$  from
(\ref{rpa}) (blue dot-dashed line) and ${\cal P}''$ from
Eq.~(\ref{new_RPA}) that describes collective response of excitons
(red solid line). The inset illustrates the relative height of the
excitonic plasmon line to the background of individual electron-hole
absorption. The position of plasmon $qv=1.006\omega$, the absorption
threshold $qv=1.081\omega$.}
\end{figure}

{\it Collective mode}. Non-zero imaginary part of the polarization
operator determines those values of $\omega$ and $q$ for which the
dissipation of external field is possible. Imaginary part of
Eq.~(\ref{pi_exact}), and as a result of Eq.~(\ref{new_RPA}), arises
in two cases: i) Imaginary $x$  corresponds to the conventional
domain, $\omega >qv$, but modified by interaction of electrons and
holes. ii) Positive $1<x<\infty$  correspond to the absorption {\it
below} the threshold and represent a non-perturbative effect which
arises from lowering the energy of  electron-hole excitations with
finite $q$ from {\it attractive} Coulomb interaction. The new
threshold of absorption $\omega = qu$ is determined from the
condition $x=1$ which for weak interactions $g\ll 1$, gives
\begin{equation}
\label{new_threshold}
 u=v[1-(g\ln{g})^2/2].
\end{equation}
Even more striking property of Eq.~(\ref{pi_exact}) is  the behavior
of its real part which describes the polarizability of an
interacting electron-hole pair. It is easy to see that the real part
is {\it positive} for $x>1$ \cite{continuation}. This change in
sign, as compared with the real part of the RPA polarizability
(\ref{rpa}), can already be traced to the first order vertex
correction (\ref{P_vertex}), which is {\it opposite} to the first
order RPA correction (\ref{P_RPA}). This fact has immediate
consequence for the {\it collective} response of excitons, given by
Eq.~(\ref{new_RPA}), which develops a pole for a certain value of
frequency $\omega(q)$. This new mode describes the propagation of
coherent oscillating charge density in a system of interacting
electron-hole pairs. The corresponding peak in the absorption has a
finite but small width which is due to the fact that the collective
mode falls within the range of decay via individual electron-hole
pairs. Its behavior is illustrated in Fig.~3 for $N=4$.

Interestingly, the width of the plasmon is  maximal around $g\sim
0.3$. For smaller values of the coupling strength the peak narrows
by virtue of the narrowing of the whole exciton domain,
Eq.~(\ref{new_threshold}). For larger values, $g\gg 1$, the peak
width decreases as $\propto g^{-2}$ while the height increases only
as $\propto g$. The position of the peak in this limit tends to the
value determined from the equation (recovered from
(\ref{pi_exact}-\ref{new_RPA}) when $x \to \infty$),
$\ln{(\frac{qv}{qv-\omega})}=N$, which yields,
\begin{equation}
\omega(q)=qv(1-e^{-N}).
\end{equation}
{\it Conclusion.} The usual RPA formalism adequately describes
dynamical charge fluctuations only in the limit $N \gg
\ln{\frac{qv}{|qv-\omega|}}$. Close to the Dirac cone we
find that the $n$-th order term in the ladder
series is larger than the corresponding term in the RPA series by a factor of $\ln^n({qv/|qv-\omega|})/N^n$. This stronger
singularity in the ladder series arises due to the long range interaction between
 electrons which move almost collinearly to the external momentum $\vec{q}$.
 The series is summed up analytically yielding
a  non-perturbative  result: the density and spin response functions
acquire non-zero imaginary part in the additional frequency range, $
q u < \omega  < q v$.  This extension is a manifestation of the
excitonic effect. The reversal of the sign of the electron
polarizability in this new domain gives rise to  a sharp plasmonic
mode which is absent in the conventional RPA. Interestingly, a
similar sign reversal of the polarizability due to interactions has
been reported to yield a  surface plasmon in a different case of a
two-dimensional Anderson insulator \cite{Raikh}.

Useful discussions with A. Chubukov, M. Raikh and O. Starykh are
gratefully acknowledged. This work was supported by DOE, 
Grant No.~DE-FG02-06ER46313.


\begin{thebibliography}{50}
\bibitem{continuation}
For $x>1$ the analytical continuation should be performed from $x\in [0,1)$ to $ (1,\infty)$. From the requirement that $\Pi''$ is negative we observe that it should yield $\sqrt{1-x^2}\to -i\sqrt{x^2-1}$. This implies that the analytical continuation must be performed in the upper half-plane of complex $x$, so that $\arcsin{x} \to \pi/2+i\ln{(x+\sqrt{x^2-1})}$.
\end{thebibliography}
\end{document}